\pdfoutput=1
\RequirePackage{ifpdf}
\ifpdf
\documentclass[pdftex]{sigma}
\else
\documentclass{sigma}
\fi

\usepackage{tikz-cd}

\numberwithin{equation}{section}

\newtheorem*{Theorem*}{Theorem}

\theoremstyle{definition}

\begin{document}

\allowdisplaybreaks

\newcommand{\arXivNumber}{2512.22114}

\renewcommand{\PaperNumber}{068}

\FirstPageHeading

\ShortArticleName{Discrete Approximations to $\operatorname U(1)$ Principal Bundles in Abelian Gauge Theory}

\ArticleName{Discrete Approximations to $\boldsymbol{\operatorname U(1)}$ Principal Bundles\\ in Abelian Gauge Theory}

\Author{Leron BORSTEN and Hyungrok KIM}

\AuthorNameForHeading{L.~Borsten and H.~Kim}
\Address{Centre for Mathematics and Theoretical Physics Research, Department of Physics, Astronomy and Mathematics, University of Hertfordshire, Hatfield, Hertfordshire AL10 9AB, UK}
\Email{\mail{l.borsten@herts.ac.uk}, \mail{h.kim2@herts.ac.uk}}

\ArticleDates{Received March 18, 2026, in final form July 06, 2026; Published online July 18, 2026}

\Abstract{A $(d+1)$-dimensional field theory with a periodic spatial dimension may be approximated by a $d$-dimensional theory with a truncated Kaluza--Klein tower of $k$ fields; as~${k\to\infty}$, one recovers the original $(d+1)$-dimensional theory. One may similarly expect that $\operatorname{U}(1)$-valued Maxwell theory may be approximated by $\mathbb Z_k$-valued gauge theory and that, as $k\to\infty$, one recovers the original Maxwell theory. However, this fails: the~${k\to\infty}$ limit of $\mathbb Z_k$-valued gauge theory is \emph{flat} Maxwell theory with no local degrees of freedom. We instead construct field theories $\mathcal T_k$ such that, with appropriate matter couplings, the $k\to\infty$ limit does recover Maxwell theory in the absence of magnetic monopoles (but with possible Wilson loops), and show that $\mathcal T_k$ can be understood as Maxwell theory with the insertion of a certain nonlocal operator that projects out principal $\operatorname{U}(1)$-bundles that do not arise from principal $\mathbb Z_k$-bundles sectors (in particular, projecting out sectors with monopole charges).}


\Classification{81T27; 70S15; 55R10; 53C05}

\section{Introduction and summary}
 It is often assumed that the continuous geometric objects characterising a physical system  are in fact limits  of  some discrete structures.
For instance, quantum field theories defined on continuum spacetime can be successfully approximated by lattice field theories \cite{Wilson:1974sk} (reviewed in \cite{Creutz:1983njd,Gattringer:2010zz,Kogut:1979wt,Montvay:1994cy,Rothe:1992nt,Smit:2002ug}).\footnote{Indeed, there have been various arguments, see, e.g., \cite{Dowker:2021zel}, that spacetime should have some fundamental discreteness, leading to approaches such as causal set theory \cite{Bombelli:1987aa, Myrheim:1978ce,Sorkin:1990bh, tHooft:1978wax} (reviewed in \cite{Dowker:2024fwa, Surya:2019ndm}).}
A related construction is dimensional deconstruction \cite{Arkani-Hamed:2001kyx}: a continuous periodic dimension $S^1$ is equivalent (via a Fourier transform) to an infinite Kaluza--Klein tower of fields, which may be truncated to obtain a discrete approximation to the continuous $S^1$ in Fourier space.

However, in one well-known case there appears to be a sharp dichotomy between the discrete and the continuous, namely for principal bundles, which are fundamental to gauge theory. Na\"ively, the Lie group $\operatorname U(1)$ may be thought of as a continuum limit $k\to\infty$ of the family of finite groups
\begin{equation*}
    \mathbb Z_k=\{\exp(2\pi\mathrm in/k)\mid n\in\mathbb Z\},
\end{equation*}
similar to lattice approximations of continuous spacetime, so that one is tempted to regard a~principal $\mathbb Z_k$-bundle as an approximation to a principal $\operatorname U(1)$-bundle. However, in fact a principal $\mathbb Z_k$-bundle differs fundamentally from a principal $\operatorname U(1)$-bundle: while generically a~principal $\operatorname U(1)$-bundles support a plethora of non-flat connections, any connection on any principal $\mathbb Z_k$-bundle is flat. This has dramatic consequences for gauge theories: the straightforward construction of a $\mathbb Z_k$-valued gauge theory must be topological \cite{Dijkgraaf:1989pz,Freed:1991bn,Tachikawa:2017gyf} and cannot approximate Maxwell theory, even for very large $k$.
Instead, the $k\to\infty$ limit of $\mathbb Z_k$-valued gauge theories only captures the \emph{flat} subsector of Maxwell theory.

This raises the question of whether it is possible to approximate full Maxwell theory in terms of discrete $\mathbb Z_k$-valued principal bundles --- more abstractly, whether there exists a theory $\tilde{\mathcal T}_k$ that complete the missing corner of the diagram:
\begin{equation*}
\begin{tikzcd}
\text{$\mathbb Z_k$ gauge theory} \rar{k\to\infty} \dar[hookrightarrow] & \text{flat Maxwell theory} \dar[hookrightarrow] \\
\tilde{\mathcal T}_k? \rar{k\rightarrow\infty} & \text{full Maxwell theory}
\end{tikzcd}
\end{equation*}

\begin{figure}
\centering
\begin{tikzpicture}[scale=1.5]
    \draw[thick] (0,0) ellipse (2cm and 1cm) node[left=3cm, below=1.5cm] {\large Maxwell theory};
    \node at (-1.25,0) {\itshape monopoles};
    \node at (0.75,0) {\itshape monopoleless sector};
    \begin{scope}
        \clip (0,0) ellipse (2cm and 1cm);
        \draw[thick] (1.5,0) ellipse (2cm and 1cm);
    \end{scope}
    \draw[thick, ->] (0.75,-1.9) -- (0.75,-1.1) node[midway, right=0.5em] {$k\to\infty$};
    \draw[thick] (1.5,-3) ellipse (2cm and 1cm) node[right=2.5cm, above=1.5cm] {\large$\mathcal T_k$};
    \begin{scope}
        \clip (1.5,-3) ellipse (2cm and 1cm);
        \draw[thick] (0,-3) ellipse (2cm and 1cm);
    \end{scope}
    \node at (0.75, -3) {\itshape admissible couplings};
    \node at (2.7, -3) {$\genfrac{}{}{0pt}0{\text{\itshape inadmissible}}{\text{\itshape couplings}}$};
\end{tikzpicture}
\caption{Correspondence between Maxwell theory and the discretised Maxwell theory $\mathcal T_k$.}\label{fig:venn}
\end{figure}
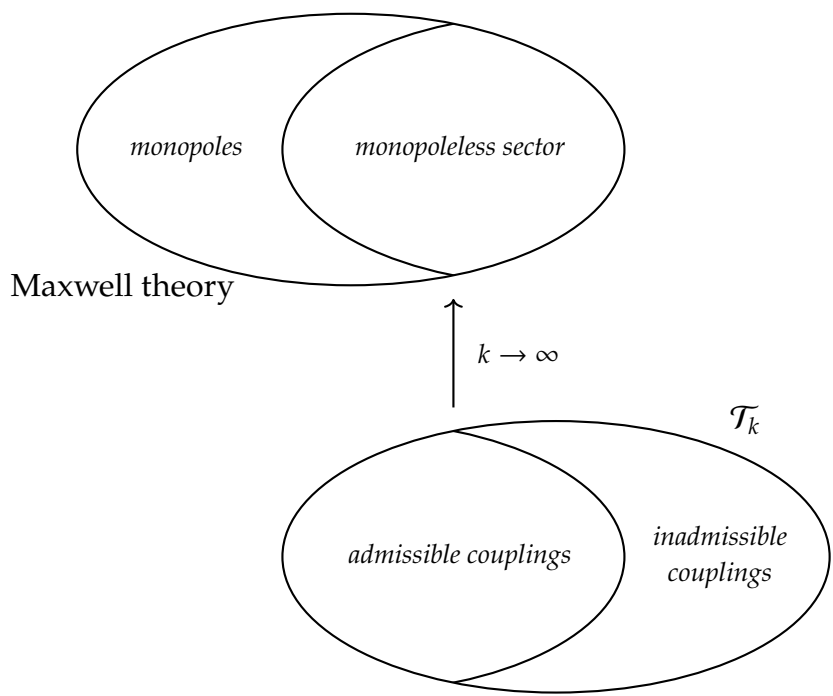
In this paper, we show that the subsector of Maxwell theory \emph{without monopoles}\footnote{More precisely,  we must restrict to $\operatorname U(1)$-bundles $P$ with vanishing first real Chern class $c_{1}(P)\in\operatorname H^2(M;\mathbb R)$ so that they are  flattenable (i.e., admit at least one flat connection). This in turn implies that the monopole fluxes, defined as $\int_\Sigma F_A$ along any closed surface $\Sigma$,  of any connection (flat or otherwise) are  vanishing. Note, however, that in the presence of torsion in $\operatorname H^2(M;\mathbb Z)$ there will nonetheless be  discrete fluxes given by  holonomies of the  flat connection, classified by  $\operatorname H^1(M;\operatorname U(1))
\cong\hom(\operatorname H_1(M;  \mathbb Z),\operatorname U(1))$.} can be approximated by a subsector of a theory that we call $\mathcal T_k$ (Figure~\ref{fig:venn}); in the $k\to\infty$ limit, this subsector of $\mathcal T_k$ tends to the monopoleless sector of Maxwell theory.
In brief,  $\mathcal T_k$ is a $\mathbb Z_k$-valued gauge theory  \big(that is, the partition function sums over principal $\mathbb Z_k$-bundles $P_{\mathbb Z_k}$\big) with a circle-valued scalar field $a$ transforming as a section of an associated bundle of $P_{\mathbb Z_k}$ and a vector gauge field $A^\sharp$ that is \emph{globally defined} (i.e.,  is a connection for the trivial $\operatorname U(1)$-bundle $\operatorname U(1)\times M$ on spacetime $M$). In addition to the $\mathbb Z_k$-valued gauge symmetry, an additional gauge transformation mixes between $a$ and $A^\sharp$. The two fields $a$ and $A^\sharp$ originate respectively from a decomposition of the Maxwell gauge field $A=A^\flat+A^\sharp$ into a flat (but not necessarily globally well-defined) part $A^\flat$ and a globally well-defined (but not necessarily flat) part $A^\sharp$ (see Section~\ref{ssec:monopoleless_maxwell}).

{\bf Future directions.}
The present analysis discretises an Abelian gauge group $\operatorname U(1)$.
A natural question is whether this may be extended to non-Abelian gauge groups. An immediate obstacle is the paucity of finite subgroups in the non-Abelian setting. For instance, for $G=\operatorname{SU}(2)$, finite subgroups admit an ADE classification \cite[Section~19]{armstrong}; the two infinite families (the cyclic and binary dihedral groups, corresponding to the $\mathrm A_n$ and $\mathrm D_n$ series respectively) are not `dense' inside $\operatorname{SU}(2)$ in the same way that $\bigcup_{k=1}^{\infty} \mathbb Z_k$ is dense inside $\operatorname U(1)$; one could instead take countably infinite subgroups of $\operatorname{SU}(2)$, but these are not truly discrete and inherit a nontrivial topology from $\operatorname{SU}(2)$.

Another direction of extension is to (Abelian) $p$-form electrodynamics, which appears more feasible. More ambitiously, one can try adjusted higher gauge theory \cite{Borsten:2024gox}, which also admits a~\v{C}ech description.

Finally, although we were motivated here by the question of discretising  Maxwell theory using $\mathbb Z_k$ gauge theory, we note that there has been much recent work on discrete gauging, including both $\mathbb Z_k$ and flat $\operatorname U(1)$ gauge theories,  in the context of generalised symmetries and the SymTFT paradigm. The literature is too vast to do justice to here, but see for example \cite{Antinucci:2024zjp,Bah:2020uev,Bhardwaj:2023kri,Bhardwaj:2017xup,Borsten:2025pvq,Borsten:2025pbx,Borsten:2025diy,Brennan:2023mmt,Cordova:2022ruw,Freed:2022iao,Gaiotto:2014kfa, Gomes:2023ahz,Kaidi:2026urc,Luo:2023ive,McGreevy:2022oyu,Nguyen:2024ikq,Santilli:2024dyz, Shao:2023gho,Sulejmanpasic:2019ytl} and references therein. It would  certainly be interesting to explore generalised symmetries in the discretised model presented here. For an initial discussion see Section~\ref{sec:checks}.

{\bf Organisation of this paper.}
In Section~\ref{sec:main}, we formulate the kinematic data of Maxwell theory in terms of \v{C}ech cocycles with respect to an open cover and discretise the data to obtain the theory $\mathcal T_k$. In Section~\ref{sec:checks}, we check that the charges, Wilson lines, and symmetries of $\mathcal T_k$ agree with those of the monopoleless sector of Maxwell theory in the limit $k\to\infty$.
In Section~\ref{sec:path_integral}, we argue that the theory $\mathcal T_k$ can be constructed ordinary Maxwell theory by inserting a nonlocal topological projection operator in the path integral.

\section{Discretisation of Maxwell theory in the \v{C}ech formulation}\label{sec:main}
\subsection{The failure of a na\"ive discretisation of Maxwell theory}

A principal $G$-bundle for $G$ a finite group admits exactly one connection, which is always flat and entirely determined by the bundle data itself \cite{Dijkgraaf:1989pz} (see also \cite{Freed:1991bn}). It is therefore immediate  that a na\"ive non-trivial discretisation of Maxwell theory must fail in the sense that the dynamical degrees of freedom are lost and cannot be recovered. To understand how one might circumvent this conclusion, it is instructive to begin with standard  Maxwell theory and see how the anticipated failure  follows upon na\"ive discretisation.  We do this here; the escape hatch that suggests itself will then be used in Section~\ref{ssec:Tk} to construct $\mathcal T_k$.

Let spacetime $M$ be an oriented manifold.
The kinematic data (i.e.,  the data over which one sums/integrates in the path integral) of Maxwell theory consists of a principal $\operatorname U(1)$-bundle \smash{$P_{\operatorname U(1)}$} and a connection $A$ on \smash{{$P_{\operatorname U(1)}$}}. If the bundle~\smash{{$P_{\operatorname U(1)}$}} is nontrivial, then $A$ is only  locally defined. In~this case, one may work with the techniques of \v{C}ech cohomology. See, for example,~\cite{Nakahara:2003nw}. Let~$M$ be covered by a sufficiently fine open cover $\{U_i\}_{i\in I}$; let each $U_i$ be connected for simplicity. Then~the principal bundle \smash{$P_{\operatorname U(1)}$} and the connection $A$ may be described as
\begin{equation*}
    g_{ij} \in \mathcal C^\infty(U_i\cap U_j,\operatorname U(1)),
    \qquad
    A_{i} \in \operatorname\Omega^1(U_i),
\end{equation*}
where $\mathcal C^\infty(X,Y)$ is the space of smooth functions $f\colon X\to Y$ and $\Omega^1(X)$ is the space of one-forms on $X$, such that
\begin{equation}\label{eq:cocycle_maxwell}
    g_{ij}g_{ji}=1,\qquad
    g_{ij}g_{jk}g_{ki}|_{U_i\cap U_j\cap U_k} = 1,\qquad
    A_{j}|_{U_i\cap U_j} = A_{i}|_{U_i\cap U_j} + \frac{\mathrm{d}(\ln g_{ij})}{2\pi\mathrm i}.
\end{equation}
The data $\{g_{ij}\}_{i,j\in I}$ form a \v{C}ech cocycle that describes a principal $\operatorname U(1)$-bundle \smash{$P_{\operatorname U(1)}$} on spacetime~$M$, while $\{A_{i}\}_{i\in I}$ specifies a connection on \smash{$P_{\operatorname U(1)}$}.

Now, if one were to replace all occurrences of $\operatorname U(1)$ by $\mathbb Z_k$, one would obtain the data
\begin{equation}\label{eq:maxwell_naive_discrete_data}
    g_{ij} \in \mathcal C^\infty(U_i,\mathbb Z_k)\cong\mathbb Z_k,\qquad
    A_{i} \in \operatorname\Omega^1(U_i),
\end{equation}
such that
\begin{equation*}
    g_{ij}g_{ji}=1,\qquad
    g_{ij}g_{jk}g_{ki} = 1,\qquad
    A_{j}|_{U_i\cap U_j} = A_{i}|_{U_i\cap U_j}.
\end{equation*}
Since $g_{ij}$ are now continuous functions with a discrete codomain $\mathbb Z_k$ (and since the domain~$U_i$ is connected), they are  \emph{constant} functions and may simply be regarded as elements of $\mathbb Z_k$. Similarly, since constant functions have zero derivatives, the term $\mathrm{d}\ln g_{ij}$ in \eqref{eq:cocycle_maxwell} vanishes, so that the patchwise defined one-forms $A_{i}\in\Omega^1(U_1)$ glue together into a one-form $A\in\Omega^1(M)$.

This na\"ive approach suffers from the following three problems:
\begin{enumerate}\itemsep=0pt
\item \emph{Decoupling between $A$ and the principal bundle.} In \eqref{eq:maxwell_naive_discrete_data}, the principal $\mathbb Z_k$-bundle $P_{\mathbb Z_k}$ defined by the cocycle $(g_{ij})_{i,j\in I}$ bears no relation to the one-form $A \in\Omega^1(M)$. Thus, it is not clear how
a $k\to\infty$ limit would relate $A$ back to a $\operatorname U(1)$-bundle obtained as a limit of~$P_{\mathbb Z_k}$.
\item \emph{All $\mathbb Z_k$-bundles are flattenable.} Another issue is whether a sequence of $\mathbb Z_k$-bundles $P_{\mathbb Z_k}$  indeed approximates an arbitrary $\operatorname U(1)$-bundle. Taking the $k\to\infty$ limit of \eqref{eq:maxwell_naive_discrete_data} would yield a principal $\operatorname U(1)$-bundle described by transition maps $g_{ij}\in\operatorname U(1)$ that are constant functions. However, a generic principal $\operatorname U(1)$-bundle is not describable in terms of transition maps that are constant; a principal $\operatorname U(1)$-bundle admits constant transition maps if and only if it is \emph{flattenable}, that is, it admits at least one flat connection.\footnote{If spacetime $M$ is trivial, so that there are no noncontractible loops around which we can have Wilson loops, then this implies that \smash{$P_{\operatorname U(1)}$} is trivial.
For $M$ not simply connected, this does not mean that \smash{$P_{\operatorname U(1)}$} is a trivial bundle, although counterexamples are nontrivial to construct.
Abstractly, constant transition maps $(g_{ij})_{i,j\in I}$ correspond to a principal bundle with fibre $\operatorname U(1)^\delta$ where $(-)^\delta$ means a group equipped with the \emph{discrete topology} (rather than the ordinary topology of Lie groups), so that such a bundle is given by the homotopy class of a map $M\to\mathrm B\operatorname U(1)^\delta$, where $\mathrm B(-)$ denotes the classifying space of a topological group. Thus a principal $\operatorname U(1)$-bundle given by the homotopy class of a map $M\to\operatorname{BU}(1)=\mathbb{CP}^\infty$ is flattenable if and only if this map factors as $M\to\operatorname{BU}(1)^\delta\to\operatorname{BU}(1)$. However, the structure of the map $\mathrm BG^\delta\to \mathrm BG$ for general $G$ is very complicated~\cite{knudson}, relating to deep statements such as the Friedlander--Milnor conjecture.} Physically, this corresponds to requiring the absence of magnetic-monopole charges since a necessary condition for a principal bundle \smash{$P_{\operatorname U(1)}$} to be flat is that, for any connection $A$ on \smash{$P_{\operatorname U(1)}$}, the magnetic monopole flux $\int_\Sigma F_A$ along any closed surface $\Sigma$ vanish;\footnote{If \smash{$P_{\operatorname U(1)}$} admits a flat connection $A^\flat$ \big(hence with curvature $F_{A^\flat}=0$\big),
    then around any closed surface $\Sigma\subset M$, the magnetic monopole charge \smash{$\int_\Sigma F_{A^\flat}$} wrapped around that surface vanishes. By Chern--Weil theory \cite{Nakahara:2003nw}, even if one choose a different, non-flat connection $A$, one always has
    \smash{$\int_\Sigma F_A=\int_\Sigma F_{A^\flat}=0$}.
}
however, nontrivial Wilson loops $\exp(\mathrm i\int_\gamma A)$ around closed loops $\gamma$ can still be present.
\item \label{item:flat_connection} \emph{All principal $\mathbb Z_k$-bundles enjoy a canonical flat connection.}
Finally, a third point of discontinuity is the fact that, while flattenable principal $\operatorname U(1)$-bundles do not have a canonical flat connection (at least if spacetime $M$ is not simply connected),
any principal $\mathbb Z_k$-bundle enjoys a canonical flat connection.\footnote{This is trivially true in the sense that the Lie algebra of the discrete group $\mathbb Z_k$ is the zero-dimensional Lie algebra $0$.}
Consequently, while associated line bundles of a flattenable principal $\operatorname U(1)$-bundle \smash{$P_{\operatorname U(1)^\delta}$} do not have canonical connections, all associated line bundles of a principal $\mathbb Z_k$-bundle $P_{\mathbb Z_k}$ always have canonical flat connections. In physics terms, this means that certain putative terms (that use this flat connection) in the action that should not be gauge-invariant in the $k\to\infty$ limit are nevertheless gauge-invariant at finite~$k$.
\end{enumerate}
We will resolve these three problems as follows:
\begin{enumerate}\itemsep=0pt
\item We decompose the connection $A$ into a flat part $A^\flat$ and a one-form part $A^\sharp$ and discretise them separately. Then $A^\flat$ retains a link to the principal bundle while $A^\sharp$ is connected to~$A^\flat$ via an additional gauge transformation.
\item We restrict to the monopoleless sector of Maxwell theory described by a flattenable principal $\operatorname U(1)$-bundle \smash{$P_{\operatorname U(1)^\delta}$}.
\item In the discretised theory $\mathcal T_k$, we restrict matter couplings to those that do not depend on the canonical flat connections on nontrivial associated bundles. (We call such couplings \emph{admissible}.)
\end{enumerate}

\subsection{Monopoleless subsector of Maxwell theory}\label{ssec:monopoleless_maxwell}
The previous discussion suggests that, in order to discretise the gauge group $\operatorname U(1)$, one should restrict to the \emph{monopoleless} subsector of Maxwell theory, where in the path integral one sums over only flattenable principal $\operatorname U(1)$-bundles. In fact, the monopoleless condition is in fact natural from the perspective of discrete-valued gauge theories. It is well known \cite{Banks:2010zn,Gaiotto:2014kfa, Maldacena:2001ss} (cf.\ the reviews \cite{Bhardwaj:2023kri, Brennan:2023mmt}) that a matter field with charge $k$ can Higgs a $\operatorname U(1)$-gauge theory down to a $\mathbb Z_k$-valued gauge theory. Globally, this mechanism can be used to realise a $\mathbb Z_k$-valued gauge theory only when the principal $\operatorname U(1)$-bundle is flattenable, i.e.,  without monopole charges since the Higgs mechanism requires a nowhere-vanishing `vacuum expectation value'\footnote{Strictly speaking, we should not speak of the vacuum expectation value of a non-gauge-invariant object such as $\Phi$, which is gauge-dependent, but only of the gauge-invariant object $|\Phi|$ \cite{Maas:2017wzi}; this is intended as a shorthand for the discussion in Appendix~\ref{ssec:higgs_argument}.} of (for instance) a charged scalar field $\Phi$ with charge $k$, that is, a globally nowhere-vanishing section of the bundle in which $\Phi$ lives, and this is only possible if $\Phi$ lives in a topologically trivial bundle, which implies that the principal $\operatorname U(1)$-bundle \smash{$P_{\operatorname U(1)}$} be flattenable. (For a detailed discussion of this point, see Appendix~\ref{ssec:higgs_argument}.)

In the monopoleless sector of Maxwell theory, any connection (i.e.,  gauge field) $A$ of a flattenable principal $\operatorname U(1)$-bundle \smash{$P_{\operatorname U(1)^\delta}$} can be (non-uniquely) put in the form
\begin{equation}\label{eq:A_split}
    A = A^\flat + A^\sharp,
\end{equation}
where $A^\flat$ is a flat connection (not necessarily defined globally) and $A^\sharp \in\Omega^1(M)$ is a one-form.\footnote{This split into a locally defined part and a one-form part also appears in, e.g., the computation of partition functions in the discussion of duality anomalies \cite{Borsten:2025phf, Donnelly:2016mlc}.} Of course, since this split is not canonical, one should introduce an additional gauge symmetry
\begin{equation*}
    A^\flat\mapsto A^\flat+\frac{\mathrm{d}\ln\alpha}{2\pi\mathrm i},\qquad
    A^\sharp\mapsto A^\sharp-\frac{\mathrm{d}\ln\alpha}{2\pi\mathrm i},
\end{equation*}
where $\alpha\in\mathcal C^\infty(M,\operatorname U(1))$ is a globally defined circle-valued function on $M$.

Furthermore, the flat connection $A^\flat$ may be represented (non-uniquely) as
\begin{equation}\label{eq:a_definition}
    A^\flat = \frac{\mathrm{d}(\ln a)}{2\pi\mathrm i},
\end{equation}
where $a$ is a section of an associated complex line bundle \smash{$L=P_{\operatorname U(1)^\delta}\times_1\mathbb C$} with respect to the canonical action of $\operatorname U(1)$ on $\mathbb C$, such that $|a|=1$ everywhere.\footnote{Equivalently, $a$ takes value in an associated circle bundle \smash{$P_{\operatorname U(1)^\delta}\times_1\mathbb S^1$}.} Of course, we still have $\operatorname U(1)$-valued gauge transformations for $A$, which we can choose to attribute entirely to $A^\flat$,
\begin{equation*}
 A^\flat \mapsto  A^\flat +\frac{\mathrm{d}\ln c}{2\pi\mathrm i},\qquad  A^\sharp \mapsto  A^\sharp\,,
 \end{equation*}
where  $c\in\mathcal C^\infty(M,\operatorname U(1))$ is a globally defined $\operatorname U(1)$-valued function.

Thus, the kinematic data describing the monopoleless sector of Maxwell theory are a flattenable $\operatorname U(1)$-bundle \smash{$P_{\operatorname U(1)}$} together with a one-form $A^\sharp \in\Omega^1(M)$ and a section $a\in\operatorname\Gamma(L)$ of the associated complex line bundle \smash{$L=P_{\operatorname U(1)^\delta}\times_1\mathbb C$} with $|a|=1$ everywhere and subject to gauge transformations
\begin{equation*}
    A^\sharp \mapsto A^\sharp- \frac{\mathrm{d}\ln\alpha}{2\pi\mathrm i},\qquad
    a \mapsto c\alpha a.
\end{equation*}

Let us examine how to couple the kinematic data \smash{$\bigl(P_{\operatorname U(1)^\delta},a,A^\sharp\bigr)$} to matter.
Since $\alpha$ was a~gauge parameter introduced to compensate for the non-uniqueness of the split $A=A^\flat+A^\sharp$, we insist that matter should not transform under $\alpha$ but only under $c$. If a complex scalar field~$\phi$ transforms as
\begin{equation*}
    \phi\mapsto c^q\phi
\end{equation*}
with charge $q\in\mathbb Z$, then $\phi$ is a section of the associated complex line bundle $P_{\operatorname U(1)^\delta}\times_q\mathbb C$. This does \emph{not} have a canonical connection, so expressions such as $\mathrm{d}\phi$ are meaningless unless $q=0$.
Instead, one could take the combination $a^{-q}\phi$, which is a true scalar field (a section of the trivial complex line bundle $M\times \mathbb C$), whose derivatives are well defined, so that one would write
\begin{equation*}
    a^q\mathrm{d}(a^{-q}\phi)
    =
    \mathrm{d} \phi
    - q\mathrm{d}(\ln a).
\end{equation*}
However, this does not couple to $A$ and thus is not gauge-invariant with respect to the transformation by $\alpha$. Compensating for this yields the gauge-covariant derivative
\begin{equation}\label{eq:covariant_derivative_maxwell}
    D^{(q)}\phi=
    a^q\mathrm{d}(a^{-q}\phi)-2\pi\mathrm iqA^\sharp\phi
    =
    (\mathrm{d}-2\pi\mathrm iqA)\phi
\end{equation}
as expected.

\subsection{Discretised Maxwell theory}\label{ssec:Tk}
We may now directly discretise the data $\bigl(P_{\operatorname U(1)},a,A^\sharp\bigr)$ by replacing $\operatorname U(1)$ with $\mathbb Z_k$. After this replacement, the discretised data are
\begin{itemize}\itemsep=0pt
\item a $\mathbb Z_k$-principal bundle $P_{\mathbb Z_k}$,
\item a one-form $A^\sharp \in\Omega^1(M)$,
\item a section $a\in\Gamma\big(P_{\mathbb Z_k}\times_1\mathbb C\big)$ of the associated complex line bundle such that $|a|=1$, where~$\times_{\bar q}$ for some \smash{$q\in\mathbb Z/k\mathbb Z$} indicates the associated complex line bundle with respect to the unitary representation\footnote{Since this action is unitary, $P_{\mathbb Z_k}\times_{\bar q}\mathbb C$ carries a canonical hermitian metric, such that the expression $|a|$ makes sense.}
\begin{equation*}
\begin{aligned}
    \mathbb Z_k&\to\operatorname{GL}(1;\mathbb C),\\
    \exp(2\pi\mathrm in/k)&\mapsto \exp(2\pi\mathrm in\bar q/k)
\end{aligned}
\end{equation*}
\end{itemize}
gauge-transforming as
\begin{equation*}
    A^\sharp \mapsto A^\sharp - \frac{\mathrm{d}\ln\alpha}{2\pi\mathrm i},\qquad
    a \mapsto c\alpha a,
\end{equation*}
where the gauge parameters are $\alpha\in\mathcal C^\infty(M,\operatorname U(1))$ and $c\in\mathcal C^\infty(M,\operatorname U(1))$.
These data define the kinematic data of the \emph{discretised Maxwell theory} $\mathcal T_k$. Note that, while the gauge parameters~$\alpha$ and~$c$ are valued in $\operatorname U(1)$, the `true' gauge symmetry group is the group of automorphisms of~$P_{\mathbb Z_k}$, which is $\operatorname{Aut}(P_{\mathbb Z_k})\cong\mathbb Z_k$; this corresponds to the subgroup $\mathbb Z_k\subset\mathcal C^\infty(M,\operatorname U(1))$.

While $a$ is a section of a nontrivial complex line bundle, now $a^k$ is a globally defined function that does not transform under $c$, and hence
\begin{equation*}
    kA = \bigl(\mathrm{d}\ln a^k\bigr)/(2\pi\mathrm i)+kA^\sharp
\end{equation*}
is a well-defined and gauge-invariant one-form.

As for Maxwell theory, matter fields are associated bundles of $P_{\mathbb Z_k}$; the $\mathbb Z_k$-representation then determines how matter should transform under $c$.
(As in Section~\ref{ssec:monopoleless_maxwell}, we require that matter transform only under $c$, not under $\alpha$.)
For instance, a complex scalar field $\phi$ with charge~$\bar q$ is a section of the associated complex line bundle $P_{\mathbb Z_k}\times_{\bar q}\mathbb C$.
It is then clear that~$\bar q$ is an integer modulo $k$ -- more abstractly, $\bar q\in\mathbb Z/k\mathbb Z$ is an~element of the Pontryagin dual group $\mathbb Z/k\mathbb Z$ of~$\mathbb Z_k$, which is isomorphic to~$\mathbb Z_k$ itself. (We however notationally distinguish between $\mathbb Z/k\mathbb Z$ and $\mathbb Z_k=\{\exp(2\pi\mathrm i\mathbb Z/k)\}$ for clarity.)

Furthermore, in light of Problem~\ref{item:flat_connection} above,  derivatives of sections of nontrivial associated bundles of $P_{\mathbb Z_k}$ are forbidden, even though they have canonical flat connections (at finite $k$). For brevity, let us say that a coupling is \emph{inadmissible} if it involves using the canonical flat connection of a nontrivial associated bundle of $P_{\mathbb Z_k}$ and \emph{admissible} otherwise.
Concretely, this means that if~${\phi\in\Gamma\bigl(P_{\mathbb Z_k}\times_{\bar q} \mathbb C\bigr)}$ is a complex scalar field with charge $\bar q\in\mathbb Z/k\mathbb Z$,
an instance of an inadmissible coupling is
\begin{equation}\label{eq:inadmissible_example}
    \mathrm{d} \phi\wedge \star\mathrm{d}\phi,
\end{equation}
since we are taking the derivative of a section $\phi$ of a nontrivial bundle. Instead, an example of an admissible coupling is
\begin{equation*}
    D^{(q)}\phi \wedge \star D^{(q)}\phi,
\end{equation*}
where
\begin{equation}\label{eq:covariant_derivative_maxwell_discrete}
    D^{(q)}\phi
    = a^q\mathrm{d}(a^{-q}\phi)-2\pi\mathrm iqA^\sharp\phi
\end{equation}
is the same expression as \eqref{eq:covariant_derivative_maxwell} for the monopoleless sector of Maxwell theory.
The expression~\eqref{eq:inadmissible_example}
is inadmissible since it involves directly taking a derivative of \smash{$\phi\in\Gamma\bigl(P_{\mathbb Z_k}\times_{\bar q} \mathbb C\bigr)$}, which uses the canonical flat connection on the bundle \smash{$P_{\mathbb Z_k}\times_{\bar q} \mathbb C$}.
Instead of writing $\mathrm{d}\phi$, we must first multiply it by an appropriate power of $a$, so that $a^{-q}\phi$ is a true scalar field, \emph{then} take the derivative, then (optionally) multiply it back by $a^q$ to land in the original line bundle, i.e.
\begin{equation*}
    a^q\mathrm{d}(a^{-q}\phi),
\end{equation*}
where $q\in\mathbb Z$ is an integer representative of $\bar q\in\mathbb Z/k\mathbb Z$.
This is, however, not gauge-invariant under transformations by $\alpha$, so that we must instead take the gauge-invariant covariant derivative \eqref{eq:covariant_derivative_maxwell_discrete}.
What happens if we change the representative $q\in\mathbb Z$ of $\bar q\in\mathbb Z/k\mathbb Z$? Suppose that $q=q'+k$. Then
\begin{equation}\label{eq:different_representatives}
    D^{(q)}\phi
    = a^{q'+k}\mathrm{d}\bigl(a^{-q'-k}\phi\bigr)-2\pi\mathrm i(q'+k)A^\sharp\phi
    =
    D^{(q')}\phi
    - 2\pi\mathrm ikA\phi,
\end{equation}
so that charges can be shifted by multiples of $k$ at the cost of producing noncanonical couplings to the gauge-invariant field $kA$. (In taking a $k\to\infty$ limit, we should hold the charges $q$ fixed, so that for sufficiently large $k$, we can always assume $|q|\ll k$.)

An action (with, say, a coupling to a complex scalar field $\phi$) can be straightforwardly written at finite $k$:
\begin{equation}\label{eq:Zkaction}
    S = \int  \biggl(-\frac1{2g^2}F\wedge \star F\biggr)
    -D^{(q)}\phi\wedge \star \bigl(D^{(q)}\phi\bigr)^*,
\end{equation}
where $F=\mathrm dA=\mathrm{d} A^\sharp\in\Omega^2(M)$
and $g$ is a coupling constant. This looks superficially similar to ordinary Maxwell theory (indeed, it is perturbatively equivalent to it), but $A^\sharp$ is no longer a connection on a principal $\operatorname U(1)$-bundle \smash{$P_{\operatorname U(1)}$} but instead a one-form. The kinematic data $\bigl(P_{\mathbb Z_k},A^\sharp,a\bigr)$ together with the action \eqref{eq:Zkaction} define the theory $\mathcal T_k$.

Then the claim is that $\mathcal T_k$
with only admissible couplings to matter tends to Maxwell theory without monopoles in the $k\to\infty$ limit when the charges $q$ of the various fields are held fixed.

\vspace{-1mm}

\section{Consistency checks}\label{sec:checks}
We must check that the discretised Maxwell theory $\mathcal T_k$ with only admissible couplings to matter in fact yields the monopoleless sector of Maxwell theory in the limit $k\to\infty$. For this, we can test whether various features of $\mathcal T_k$, such as the perturbative theory, its matter charges, and symmetries, approximate those of the monopoleless sector of Maxwell theory.

\subsection{Perturbative equivalence}
One immediate check is whether the number of local degrees of freedom match between the monopoleless sector of Maxwell theory and $\mathcal T_k$. Maxwell theory enjoys $d-2$ local degrees of freedom in $d$ spacetime dimensions transforming as a vector field as does $\mathcal T_k$ (and unlike pure $\mathbb Z_k$-valued gauge theory, which lacks any local degrees of freedom, being purely topological); in fact, the two theories are equivalent perturbatively, with the exactly the same set of Feynman diagrams, as follows directly from the action \eqref{eq:Zkaction}.

\subsection{Matter charges}
In Maxwell theory, electric charges correspond to the irreducible representations of $\operatorname U(1)$, which are all one-dimensional and in canonical bijection with $\mathbb Z$ (the Pontryagin dual group of $\operatorname U(1)$, see, e.g., \cite{Bhardwaj:2023kri}). Full Maxwell theory also has magnetic charges, but these are absent by definition in the monopoleless sector of Maxwell theory.

In the discretised Maxwell theory $\mathcal T_k$, charges correspond to different associated bundles of~$P_{\mathbb Z_k}$, which are labelled by elements $\bar q\in\mathbb Z/k\mathbb Z$.
For each $\bar q$, there are inequivalent ways of writing down a covariant derivative $D^{(q)}$ labelled by a choice of a representative $\bar q=q+k\mathbb Z$. However, such terms differ by a noncanonical coupling proportional to $k$ as seen in \eqref{eq:different_representatives},
and there is no corresponding conservation law for $q\in\mathbb Z$ (rather than $\bar q\in\mathbb Z/k\mathbb Z$) since the gauge symmetry is valued in $\mathbb Z_k$ (the Pontryagin dual to $\mathbb Z/k\mathbb Z$), not in $\operatorname U(1)$ (the Pontryagin dual to~$\mathbb Z$).\looseness=-1

However, as $k$ tends to infinity, the set of charges $\mathbb Z/k\mathbb Z$ of $\mathcal T_k$ approximates the set of charges~$\mathbb Z$ of Maxwell theory.

\subsection{Wilson loops}
The classification of charges similarly extends to the classification of Wilson loops.
For an~ordinary gauge theory with a compact gauge group $G$, along any closed loop $\gamma$ in spacetime $M$,
one can construct a Wilson loop for any (without loss of generality irreducible) finite-dimensional representation $\rho$ of $G$, representing the holonomy of a particle transforming as $\rho$ under the gauge symmetry; in general, the possible Wilson loops are in bijection with the possible charges.

Specialising to the case $G=\operatorname U(1)$, in Maxwell theory, along any closed loop $\gamma$ in spacetime~$M$, and for any integer $q\in\mathbb Z$ labelling a (one-dimensional) complex irreducible representation of~$\operatorname U(1)$, we can construct the Wilson loop{\samepage \vspace{-1mm}
\begin{equation*}
    W_\gamma[A]=\exp\oint_\gamma 2\pi \mathrm iqA\vspace{-1mm}
\end{equation*}
that is invariant under $\operatorname U(1)$ gauge transformations.}

In the case of the discretised Maxwell theory $\mathcal T_k$, similarly, one has Wilson loops
\begin{equation*}
    W_\gamma[A]=\exp\oint_\gamma q(2\pi\mathrm iA)
    =\biggl(\exp\oint_\gamma 2\pi\mathrm iqA^\sharp\biggr)
    \biggl(\exp\oint_\gamma q\,\mathrm d\ln a\biggr)
\end{equation*}
for $q\in\mathbb Z$.\footnote{The latter factor does not vanish since $\mathrm d\ln a$ is closed but not exact: $\ln a$ is not globally defined.} Note that each of the two factors $\exp\oint_\gamma 2\pi\mathrm iqA^\sharp$ and $\exp\oint_\gamma q\,\mathrm d\ln a$ are separately invariant under transformation by $c\in\mathbb Z_k$, but not under gauge transformation by $\alpha\in\mathcal C^\infty(M,\operatorname U(1))$.
Thus, gauge-invariant Wilson loops of $\mathcal T_k$ are labelled by an integer $q\in\mathbb Z$ in agreement with those of Maxwell theory.

\subsection{Higher-form symmetries}
Maxwell theory has a one-form electric symmetry whose Noether current is $\star F$ as well as a $(d-3)$-form magnetic symmetry whose Noether current is $F$ \cite{Gaiotto:2014kfa}.
If one restricts to the sector with no magnetic monopoles, however, the magnetic $(d-3)$-form symmetry becomes trivial since nothing is charged under it (equivalently, it does not survive in the cohomology \cite{Borsten:2025pbx});
therefore, the monopoleless sector of Maxwell theory only has a $\operatorname U(1)$-valued one-form symmetry. In addition, since $g_{ij}$ can be now made to be constant, the cocycle $(g_{ij})_{i,j\in I}$ defines a $\operatorname U(1)$-valued \v{C}ech one-cocycle corresponding to a $(d-2)$-form symmetry. Assuming $M$ compact (and oriented), then Poincar\'e duality then produces a corresponding $(d-1)$-cocycle corresponding to a $\operatorname U(1)$-valued one-form symmetry.

The discretised Maxwell theory $\mathcal T_k$ has a one-form symmetry given by $\star\mathrm dA$; in addition, from the $\mathbb Z_k$ gauge symmetry it has a $\mathbb Z_k$-valued one-form symmetry and a $\mathbb Z_k$-valued $(d-2)$-form symmetry, given by the \v{C}ech one-cocycle $(g_{ij})_{i,j\in I}$ and its Poincar\'e dual, respectively \cite{Gaiotto:2014kfa} (as reviewed in, e.g., \cite{Bhardwaj:2023kri}). Thus, in the limit $k\to\infty$ where $\mathbb Z_k$ approximates $\operatorname U(1)$, the sets of symmetries of $\mathcal T_k$ agree with those of the monopoleless sector of Maxwell theory.

\section[Discretised Maxwell theory as ordinary Maxwell theory with nonlocal operator insertion]{Discretised Maxwell theory as ordinary Maxwell theory\\ with nonlocal operator insertion}\label{sec:path_integral}
In this section, we approach the discretised Maxwell theory $\mathcal T_k$ in a different way.
We show that~$\mathcal T_k$ may be seen as ordinary Maxwell theory with the insertion of a nonlocal operator in the path integral that projects down to only $\mathbb Z_k$-principal bundles and which then breaks the gauge symmetry down to $\mathbb Z_k$.

The path integral of Maxwell theory on a smooth oriented spacetime $M$ sums over all connections $A$ for all principal $\operatorname U(1)$-bundles over $M$:
\begin{equation*}
    Z(M) = \sum_{P_{\operatorname U(1)}\in[M,\operatorname{BU}(1)]}\int\mathrm DA\,\exp(\mathrm iS[A]),
\end{equation*}
where $[M,\operatorname{BU}(1)]$ is the discrete set of all principal $\operatorname U(1)$-bundles on $M$ up to topological isomorphism.\footnote{Or, equivalently, the set of homotopy classes of continuous maps from $M$ to the classifying space $\operatorname{BU}(1)=\mathbb{CP}^\infty$ (infinite-dimensional complex projective space) of the Lie group $\operatorname U(1)$.}

One can insert various operators into this path integral.
Consider the topological nonlocal operator $O$, whose value is\footnote{Here, `arises from a $\mathbb Z_k$-principal bundle' means that the transition maps $g_{ij}$ can be taken to be constant and valued in $\mathbb Z_k$; equivalently, the continuous map $f_{\operatorname U(1)}\colon M\to\operatorname{BU}(1)$ whose homotopy class defines the principal $\operatorname U(1)$-bundle \smash{$P_{\operatorname U(1)}$}
    in fact factorises as
    $
        f_{\operatorname U(1)} = \mathrm B\iota \circ f_{\mathbb Z_k}
    $
    where $\mathrm B\iota \colon\mathrm B\mathbb Z_k\to\operatorname{BU}(1)$ is induced by the group inclusion $\iota\colon\mathbb Z_k\to\operatorname U(1)$.
    Non-isomorphic principal $\mathbb Z_k$-bundles can give rise to isomorphic $\operatorname U(1)$-bundles: for example, on a~circle~$\mathbb S^1$, there are $k$ inequivalent $\mathbb Z_k$-bundles (classified by group homomorphisms $\mathbb Z\cong\pi_1(\mathbb S^1)\to\mathbb Z_k$), but these all give rise to the trivial $\operatorname U(1)$-bundle on $\mathbb S^1$ up to isomorphisms of principal $\operatorname U(1)$-bundles \big(in fact, there is only one $\operatorname U(1)$-bundle on $\mathbb S^1$ since \smash{$\operatorname H^2\bigl(\mathbb S^1\big)=0$} is trivial\big). The number of  inequivalent  $\mathbb Z_k$-bundles that give rise to a fixed $\operatorname U(1)$-bundle  \smash{$P_{\operatorname U(1)}$}  is \smash{$k^{b_1}$}, where \smash{$b_1=\dim\operatorname H^1(M; \mathbb Z)$}. This follows from reducing the integral cohomology modulo~$k$ in the morphism  $\operatorname H^1(M; \mathbb Z) \to (\mathbb Z_k)^{b_1}  \subset\operatorname H^1(M; \mathbb Z_k) = \hom(\operatorname H_1(M; \mathbb Z), \mathbb Z_k) = (\mathbb Z_k)^{b_1} \oplus \hom(\operatorname{Tors}(\operatorname H_1(M; \mathbb Z)), \mathbb Z_k)$, where $\operatorname{Tors}(-)$ denotes the torsion component of a finitely generated Abelian group. \big(Note that the choices in  $\hom(\operatorname{Tors}(\operatorname H_1(M; \mathbb Z)), \mathbb Z_k)$ will give rise to inequivalent $\operatorname U(1)$-bundles, whereas we are only counting those $\mathbb Z_k$-bundles yielding a fixed \smash{$P_{\operatorname U(1)}$}.\big)}
\begin{equation*}
    O [P_{\operatorname U(1)},A ] =\# \{\text{isomorphism classes of $\mathbb Z_k$-bundles giving rise to \smash{$P_{\operatorname U(1)}$}} \}.
\end{equation*}
\big(In particular, if \smash{$P_{\operatorname U(1)}$} does not arise from any principal $\mathbb Z_k$-bundle, then $O\bigl[P_{\operatorname U(1)},A\bigr]=0$; this is the case whenever \smash{$P_{\operatorname U(1)}$} is not flattenable.\big)
The operator $O$ is a topological operator since it does not depend on the connection $A$; it is a nonlocal operator because it is not the integral of a local quantity (unlike operators for Chern numbers, which are expressible as integrals of quantities constructed from the curvature of $A$). To give a concrete example of such an operator, recall that a flattenable $\operatorname U(1)$-bundle \smash{$P_{\operatorname U(1)}$} can arise from a $\mathbb Z_k$-bundle if and only if $kc_1\bigl(P_{\operatorname U(1)}\bigr)=0$.  This condition  can  be imposed by inserting a Dirac delta $\delta\bigl(kc_1\bigl(P_{\operatorname U(1)}\bigr)\bigr)$ into the partition function via a $BF$-type term
\begin{equation*}
\int\mathrm D B
\exp\biggl(
\frac{\mathrm i k}{2\pi}\int_M B\wedge F_A
\biggr),
\end{equation*}
where $B$ is a  $(d-2)$-form Lagrange multiplier field,  imposing
\begin{equation*}
k\frac{F_A}{2\pi}=0,
\end{equation*}
which implies \smash{$k c_1(P_{\operatorname U(1)})=0$}. However, each \smash{$P_{\operatorname U(1)}$} surviving this condition could have arisen from \smash{$k^{b_1(M)}$} inequivalent $\mathbb{Z}_k$-bundles, where $b_1(M)$ is the first Betti number of $M$, so we should include the multiplicity factor:
\begin{equation*}
O[P_{\operatorname U(1)}, A] =k^{b_1(M)} \int\mathrm D B
\exp\biggl(
\frac{\mathrm i k}{2\pi}\int_M B\wedge F_A
\biggr).
\end{equation*}

If one considers the path integral with $O$ inserted as
\begin{equation*}
    Z(M) = \sum_{P_{\operatorname U(1)}\in[M,\operatorname{BU}(1)]}\int\mathrm DA O\bigl[P_{\operatorname U(1)},A\bigr]\exp(\mathrm iS[A]),
\end{equation*}
the sum then restricts to a sum over those $\operatorname U(1)$-bundles that arise from $\mathbb Z_k$-bundles; by construction these are always flattenable so that one can decompose $A$ into $a$ and $A^\sharp$ as in \eqref{eq:A_split} and \eqref{eq:a_definition}, and then the Maxwell action then reduces to that of $\mathcal T_k$ theory.

To summarise:  a  continuous Maxwell theory on a $\operatorname U(1)$-bundle \smash{$P_{\operatorname U(1)}$} will restrict to a  $\mathcal T_k$ if \smash{$P_{\operatorname U(1)}$} is  flattenable and \smash{$kc_1\bigl(P_{\operatorname U(1)}\bigr)=0$}.  Note, however, that  the definition of  a $\mathcal T_k$ theory is independent of any continuous  Maxwell theory origin. When a continuous Maxwell theory restricts to a  $\mathcal T_k$, there exist  transition functions that give cocycles in $\operatorname H^1(M; \mathbb{Z}_k)$ along with the corresponding  discrete holonomies valued in $\mathbb{Z}_k$.

\appendix
\section{Monopolelessness from the Higgs mechanism}\label{ssec:higgs_argument}
Let us justify the claim made in Section~\ref{ssec:monopoleless_maxwell} that the monopoleless condition appears naturally in Higgsing a $\operatorname U(1)$-gauge theory down to a $\mathbb Z_k$-valued gauge theory using matter with charge $k$.

Consider the Lagrangian of a complex scalar field $\Phi$ of charge $k$ coupled to Maxwell theory and subject to a Mexican-hat potential $V(|\Phi|)$ on a compact, oriented, and connected (but not necessarily simply connected) spacetime $M$. If spacetime $M$ is not simply connected, then one can put twisted boundary conditions for a complex scalar field on $M$; such boundary conditions are then classified by group homomorphisms
\begin{equation*}
    \chi\colon\ \pi_1(M)\to\operatorname{GL}(1;\mathbb C),
\end{equation*}
where $\pi_1(M)$ is the fundamental group of $M$, specifying the compatible holonomies of $\Phi$ across each non-contractible cocycle.
Then $\Phi$ is a section of the complex bundle
\begin{equation}\label{eq:higgs_line_bundle}
    L = L_\chi \otimes \big(P_{\operatorname U(1)}\times_k \mathbb C\big),
\end{equation}
where \smash{$P_{\operatorname U(1)}$} is the principal $\operatorname U(1)$-bundle of Maxwell theory, and $L_\chi$ is the flat complex line bundle on $M$ whose sections are uncharged scalar fields obeying the twisted boundary conditions given by $\chi$. Note that $L$ admits a Hermitian metric (since the one-dimensional representation of~$\operatorname U(1)$ with charge $k$ is unitary), so it makes sense to speak of $|\Phi|$, which is merely a function $|\Phi|\colon M\to\mathbb R_{\ge0}$ rather than the section of a nontrivial fibre bundle.

We then carry out the usual discussion of the Higgs mechanism. The Lagrangian density is
\begin{equation*}
    \mathcal L
    =
        -\frac1{2g^2}F\wedge \star F
        -
        (D\Phi)\wedge \star
        (D\Phi)
        -V(|\Phi|),
\end{equation*}
where
\begin{equation*}
    D = \mathrm d  + 2\pi\mathrm ikA
\end{equation*}
is the covariant derivative.
Supposing that $V(|\Phi|)$ has a minimum at $|\Phi|=v$ (and at no other value of $|\Phi|$), we may uniquely decompose (as long as $|\Phi(x)|$ vanishes nowhere)
\begin{equation*}
    \Phi(x) = \frac1{\sqrt2}(v+\phi(x))\exp(\mathrm i\theta(x)),
\end{equation*}
where $\phi(x)$ is a real scalar field, and $\theta$ is a section of the circle bundle associated to $L$. Define
\begin{equation*}
    B = A+\mathrm d \theta,\qquad G = \mathrm d B.
\end{equation*}
In this parameterisation, the Lagrangian density becomes
\begin{equation*}
    \mathcal L =
        -\frac1{2g^2}G\wedge \star G
        -\frac12\mathrm d\phi\wedge \star \mathrm d\phi
        -V(\phi+v)
        -\frac12
        (v+\phi)^2
        (\mathrm d\theta+kA)\wedge \star
        (\mathrm d\theta+kA).
\end{equation*}
Now, $\phi$ takes the form of a massive scalar with squared mass given by the second derivative of~$V$.
If we can gauge-fix $\theta$ to some specific value, then this would be equivalent to the Proca action for a massive vector field coupled to the massive scalar $\phi$. This requires that $\theta$ be globally well defined, i.e.,  that the circle bundle associated to $L$ have a global section.
This is only possible if $L$ is topologically the trivial complex line bundle.
This, in turn, is only possible if \smash{$P_{\operatorname U(1)}$} is flattenable. (In that case, one can take \smash{$L_\chi = P_{\operatorname U(1)}\times_{-k}\mathbb C$}
in \eqref{eq:higgs_line_bundle} since \smash{$ P_{\operatorname U(1)}\times_{-k}\mathbb C$} admits a~flat connection.) Otherwise, $\theta$ only admits local sections, and the global physics differs from a~Proca field even though the local physics is the same.

Since the local physics agrees with that of a Proca field, the spectrum is gapped, and in the deep infrared one has at most a topological theory. The argument of \cite{Banks:2010zn} requires passing to the deep infrared and keeping only the term $A \wedge \star \mathrm d \theta$; if $\theta$ can be dualised into a $(d-2)$-form, this would yield the action for a $BF$ formulation of $\mathbb Z_k$ gauge theory. For this to work, $\theta$ needs to take values in the trivial circle bundle, i.e.,  \smash{$P_{\operatorname U(1)}$} must be flattenable. Equivalently, the equation of motion for $\theta$ then requires $\mathrm dA=0$, so that $A$ defines a flat connection for \smash{$P_{\operatorname U(1)}$}, which must be therefore flattenable.

\subsection*{Acknowledgements}
Leron Borsten is grateful for the hospitality of the Theoretical Physics group, Blackett Laboratory, Imperial College London.
The authors thank the anonymous referees, whose helpful comments improved the manuscript.

\pdfbookmark[1]{References}{ref}
\LastPageEnding

\end{document}